# An ICT based Solution for Virtual Garment Fitting for Online Market Place: A Review of Related Literature

Hashini Gunatilake[*], Dulaji Hidellaarachchi, Sandra Perera
School of Computing
University of Colombo
Colombo, Sri Lanka
[*]Email: hashini17s1 [AT] gmail.com

Damitha Sandaruwan, Maheshya Weerasinghe
School of Computing
University of Colombo
Colombo, Sri Lanka

*Abstract*— **In this paper, we describe various technologies that are being used in virtual garment fitting and simulation. There, we have focused about the usage of anthropometry in clothing industry and avatar generation of virtual garment fitting. Most commonly used technologies for avatar generation in virtual environment have been discussed in this paper such as generic body model and laser scanning. Moreover, this paper includes the real-time tracking technologies used in virtual garment fitting like markers and depth cameras in various related researches as well as how the virtual cloth generation and simulation carried out in the related researches. Apart from these, virtual clothing methods such as geometrical, physical and hybrid based models were also discussed in this paper. As ease allowance has a major impact on virtual cloth fitting, it is also considered in this paper related to similar researches. Within this paper, all the above mentioned areas were described thoroughly while stating the existing gap of the virtual garment fitting in online marketplaces.**

## I. Introduction

Nowadays people purchase garments online due to their busy lifestyles. It is visible that these online purchases have a considerable return rate. The main reason behind this situation is the size mismatch. There are many accurate computationally intensive 3D solutions to check the fitness of a garment with the human body. The problem associated with these 3D solutions is that we cannot use them to obtain a real time result for the online market place since they require human intervention and 3D mesh models of users and apparels. Further there will be large number of users who will be accessing the system concurrently to check their fitness with a dress. Even though the available 3D solutions will provide an accurate solution, the time consumed by the 3D solutions to deliver the solution will be very high. An online marketplace will require a quick solution and the available 3D solutions will not be appropriate to fulfill this requirement. Furthermore, according to a study done by D. Kim and K. LaBat (2013), considering the consumer experience in using 3D virtual garment simulation technology [1], most of the participants thought 3D virtual garment simulation would be a good starting point for judging fit, but there were participants with different opinions. They have realized that there will be privacy issues while using these 3D solutions, and the lack of the availability of technology was also an issue. Further the discomfort of viewing one's own body scan was a huge issue while using these 3D virtual simulation technologies. Therefore there is a need for a lightweight solution which will support the decision making of the end user in order to select the properly fitted dress by considering human body properties and garment properties.

## II. Usage of Anthropometry in Clothing Industry

According to Pheasant and Haslegrave (2006) anthropometry is a branch of human science which deals with the measurements of the human body in terms of size, shape, mobility, flexibility and working capacity [2]. It is observable that human body dimensions and proportions vary from one to another which triggers the necessity of understanding and analyzing the variances of the human body dimensions as well as the relationship among those. Pheasant and Haslegrave (2006) have presented correct standing and sitting postures in which the extraction of human body measurements could be obtained more accurately [2].

However, inception of massive cloth production for general public has triggered standardization of clothing sizes according to a clothing pattern grading system that was introduced by experts in fashion industry. According to Schofield and LaBat (2005), aforementioned pattern grading method presents a standard way of making adjustments to cloth design patterns with reference to predefined set of human body sizes [3]. This pattern grading system has a strong correlation with anthropometry. According to Huang (2012), the apparel industry uses ASTM D5219-09 and ISO 8559:1989 standards for taking anthropometric measurements and locating anatomical landmarks on human bodies [4]. These standards are used to locate body features on the parametric human models for which 3D garment should align with.

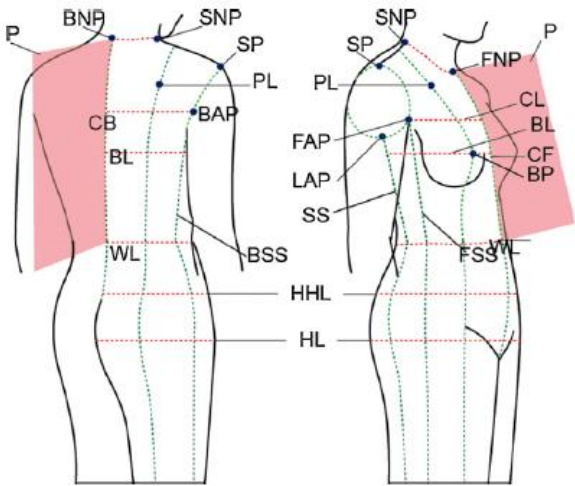

Figure 1. Anthropometric body features (Waist line (WL), Bust point (BP), Bust line (BL), Hip line (HL), High hip line (HHL), Center front (CF), Center back (CB), Front neck point (FNP), Side neck point (SNP), Back neck point (BNP), Shoulder point (SP), Chest line (CL), Front arm point (FAP), Back arm point (BAP), Lowest armhole point (LAP, Side seam (SS), Princess line (PL), Front side seam (FSS) and Back side seam (BSS) ) [4]

According to Chen (2008), there are eight body measurements required to be obtained for a normal trouser [5], and these parameters can be classified into two classes as follows: (1) Vertical type. Waist to hip, out leg, curved front body rise. (2) Girth type. Thigh girth, waist girth, half back waist, hip size, half back hip.

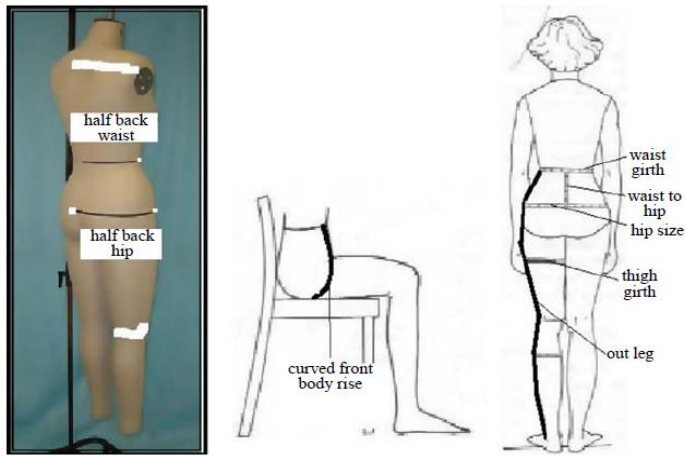

Figure 2. Relevant body measurements related to the gluteal region and the trouser of normal size [5]

### III. AVATAR GENERATION FOR VIRTUAL GARMENT FITTING

Generation of avatars represents human body in various virtual fitting applications where human body measurements are considered as one of the main inputs which can be entered manually to generate an avatar. This is commonly used when using web based applications and apart from measurements of human body can also be extracted from numerous measuring and capturing technologies such as laser scanning, multi-camera capturing, motion detection and etc.

#### A. Use of Generic body models

One popular method of representing virtual human body construction and animation is using generic models [6]. There, the creation of face, body skeleton and body surface have been considered using the generic models. This avatar generation approach starts from default virtual human templates that include shape and animation structures. As one way, photographs have been used as inputs, where front and side views of the face and front, side and back views of the body were taken. This method is a feature-based modeling of animatable body. Following several steps the generic model is animated and performs the animation with skin deformation. Then a generic body is gained with seamless surface and real-time skin deformation capability whenever skeleton joints are animated with given animation parameters.

In another way, certain measurements of human body have been used to adjust the generic models to generate the avatar. There, they have taken eight length measurements that has to be entered by the user and based on that particular segments of avatar are adapted and others are done using interpolations. Those measurements are as follows in Table 1.

TABLE I. EIGHT BASIC MEASUREMENTS TO FORM AN AVATAR [6]

| Body measurement | Definition |
|---|---|
| Stature | Vertical distance between the crown of the head and the ground |
| Crotch length | The vertical distance between the crotch level at center of body and the ground |
| Arm length | The distance from the armscye shoulder line intersection(acromion) over the elbow to the far end of the prominent wrist bone(ulna) in line with small finger |
| Neck girth | The girth of the neck-base |
| Chest/Bust girth | Maximum circumference of the trunk measured at bust/chest height |
| Underbust girth | Horizontal girth of the body immediately below the breasts |
| Waist girth | Horizontal girth at waist height |
| Hip girth | Horizontal girth of the trunk measured at hip height |

Moreover, by creating generic human body model for standard five types of sizes (Extra Small, Small, Medium, Large, Extra-large) for each gender, and then taking the measurements (defined primary measurements, Table 2) an avatar generat

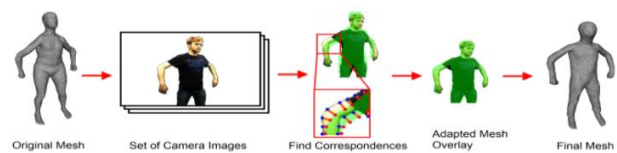

ion is carried out [7].

TABLE 2. PRIMARY MEASUREMENTS USED TO GENERATE FEMALE HUMAN BODY MODEL [7]

| | Female Standard Body Sizes | XS/34 | S/36 | M/38 | L/40 | XL/42 |
|---|---|---|---|---|---|---|
| | Primary Measurements | | | | | |
| 1 | Height | 168 | 168 | 168 | 168 | 168 |
| 2 | Bust girth | 80 | 84 | 88 | 92 | 96 |
| 3 | Under-bust girth | 71 | 74 | 77 | 80 | 84 |
| 4 | Waist girth | 65 | 68 | 72 | 76 | 80 |
| 5 | Hip girth | 90 | 94 | 97 | 100 | 103 |
| 6 | Inside leg length | 78.3 | 78.3 | 78.1 | 77.9 | 77.7 |
| 7 | Arm length | 59.6 | 59.8 | 60 | 60.2 | 60.4 |
| 8 | Neck-base girth | 34.8 | 35.4 | 36 | 36.6 | 37.2 |

*B. Use of Laser Scanning*

The creation of three-dimensional avatar can also be done using laser scanning [8]. Using 3D laser scanner, customers body surface will be scanned within a few seconds and produces a three-dimensional point cloud. Then a mesh is generated out of these scanned point clouds and textures and acts as the basis for the virtual avatar. By applying skin deformation methods to this created basis of virtual avatar, the finalized avatar is capable of generating simple animations as well.

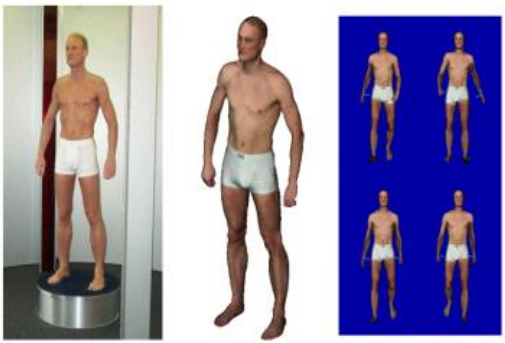

Figure 3. Different stages creating an individual customer avatar : Scanning the customer's body surface (left). Static avatar (middle). Dynamic avatar (right) [8]

*C. Use of Optical Tracking*

Using of optical tracking with commodity cameras is suggested in order to to scan a person and then the scanned body mesh will be used as an avatar later [10]. To capture all sides of a person's body measurements, multi-camera setup with ten cameras will be used and comparing to other methods, this is used because of its fast processing. With the pose estimation algorithms [11] the images are converted to a depth map and by using deformations and based on the images, the mesh is matched in order to resemble the proportions of the captured person which can be shown in figure 4 illustrated from [10].

Figure 4. The transformations that are applied on the original mesh based on the multiple camera images [10].

More natural look of avatar generation can be achieved using face detection algorithm on a photo of a user. A further adjustment such as skin color is done according to the face [9].

Though there are different ways of generating an avatar, most of these are unable to use in real-time applications as it requires huge computational power. As a result, there is a need of an avatar generation method which requires low computational power that can be used in online market place.

## IV. REAL-TIME TRACKING TECHNOLOGIES FOR VIRTUAL GARMENT FITTING

These technologies can be categorized into two parts as marker-based technologies and markerless technologies. Marker-based approach uses different techniques. Fiducial markers and optical markers are such techniques that come under this approach. Depth camera technique is known as a sophisticated markerless technology which enables motion capturing and full body depth detection.

*A. Markers*

Marker-based approach can be used as one of the methods to track shape of the body by placing these markers on predefined positions of the surface of body. These are known as fiducial markers since placement of markers are done in accordance to the comparison of placements of other markers.

In a study that presented an Augmented Reality application called ARDressCode has applied concept of markers for their study [12]. Three cameras have been setup to capture shopper's body, and using of Augmented Reality Toolkit, the images captured by the camera are analyzed in order to place a set of predefined markers. As final step, 3D model of garment is positioned on body considering markers. In another study done in 2008 [13], researchers have applied markers on joints of user using an algorithm that detects accurate positions of body. Finally a 2D garment is positioned on the body of the user based on markers.

With the introduction of markerless tracking system it was identified that marker-based approach is an error-prone method as real-time positioning of markers were less accurate.

*B. Depth Cameras*

In a study done in Japan, 3D body shapes of participants were obtained using depth cameras in order to overlay clothing images that is with similar body shape [14]. The subjects are captured with depth cameras, resulting single-shot depth images. This single-shot depth image is used to obtain 3D body model of subject with the application of depth data. Once single-shot depth image is rendered, shape of the body model is estimated using depth data. Finally the most suitable clothing image with similar shape is selected from clothing database.

This method was able to produce more accurate garment fitting for the users, but with the high cost of large databases and equipment, the usage of depth cameras is inappropriate for an online marketplace.

### V. VIRTUAL CLOTH GENERATION AND SIMULATION

Generation and simulation of virtual cloth can be done in various ways such as creating a 3D mesh to represent cloth, scanning an existing cloth using laser scanning techniques or using 2D planes with considering the texture of the fabric. From a research done in Graz University of Technology [10], they have generated the garment items virtually based on a multi-camera setup. With the assistance of this set up, a person will be captured who is wearing the actual cloth and the cloth should be separated from the body which can be done both manually or using chroma key technique. In another method which is used for cloth generation [6][15], described in two research papers which was published in 2001 and 2003 respectively, first the outline of the cloth is drawn and placed on the body. After competing series of steps finally, the garment is adapted automatically according to the shape of the model. With a different technique that provides a realistic visualization of clothes is presented in a research [8] it have been mentioned that a finished results of cloth generation and simulation should give the real feeling of material properties to the users. In this method, clothes are created using 2D CAD geometry models and these models are typically utilized in the fashion industry.

Another research which was conducted in University of Geneva [7], they have used 3D garment simulator namely MIRACloth that was developed in University of Geneva. In this research they have conducted the cloth simulation in two stages, garment assembly stage and garment animation stage.

### VI. VIRTUAL CLOTHING IN COMPUTER GRAPHICS

Virtual clothing supports the reproduction of physical behaviors and the visual features of textile objects in computer simulated virtual reality [16]. Virtual clothing can be classified into three main categories such as geometrical based, physical based and hybrid virtual clothing methods. Distinction among these categories is identified based on the core technique used for formulating shape of the cloth or for driving its deformation.

#### A. Geometrical Based Virtual Clothing Methods

These techniques mainly concentrate on the properties related to appearance particularly folds and wrinkles which can be represented by geometrical equations. Physical properties of cloth will not be examined through these methods. Geometrical based virtual clothing methods require a substantial amount of user intervention which can be considered as one of its limitations.

The roots of geometrical virtual clothing methods spread back into 1986. Weil (1986) was able to introduce a way of modelling for three dimensional (3D) hanging cloth material [17]. Catenary curves between hanging points of cloth were used to induce shape of the cloth. This method can be used only to generate the hanging cloth and it cannot be used to generate a more complex cloth shape. T. Agui and Nakajima (1990) presented a way of modelling a sleeve on a bent arm [18]. According to their observation, consequence of differences in curvature between inner and outer part of the sleeve has caused formation of folds. This method focuses mainly the simulation of bent sleeve. Hinds and McCartney (1990) aimed at automation of manufacturing garments [19]. In this method upper body of a mannequin was digitized to obtain shape of human body and numerous three dimensional (3D) panels were used to represent a garment. Afterwards Hinds et al. (1991) proposed a way of translating three dimensional(3D) panels into two dimensional (2D) patterns using method of Calladine (1986) [20] [21]. Miller et al. (1991) proposed a new approach to the issue of producing a simple topologically-closed geometric model from a point-sampled volume data set [22]. A simple geometry as an initial topologically closed object such as a sphere or a cube was proposed through their research. Then the process of expanding this simple object in order to fit the object within a volume has been conducted. An extension of this method into cloth modelling was proposed by Thomas Stumpp (2008) [23]. Linear time complexity was reached considering number of mesh vertices through their method. Physical property of cloth can be affected by the topology of cloth mesh as physical property of the cloth is correlated to the size of clusters. Decaudin et al. (2006) introduced a method to create visually realistic clothes by wrapping developable surfaces around the character body in a natural manner [24]. The sewing patterns from 3D cloth model were provided by flattening the developable surfaces. A three dimensional (3D) surfaces were generated around character body using a sketch based interface and manually added the seam lines on these surfaces directly. A feature based method for the construction of 3D cloth from 2D sketches was proposed by Wang et al. (2003) by considering predefined human body features as base [25]. Tailoring rules in fashion industry were used to define features of human body based on the profiles of different body parts. 3D cloth templates are pre-constructed considering each type of cloth. This method can generate only simple cloth meshes which needed to be processed further to complete a detailed cloth. An interactive way for putting and manipulating clothes on a 3D model was introduced by Igarashi & Hughes (2002) [26]. 2D patterns are used to generate 3D cloth in this method and this can be used to model simple style cloth without folding and also because of the computational cost cloth-cloth collision is disregarded while calculating surface constraints of 3D cloth mesh. An approach for designing 3D cloth directly on a 3D mannequin model was introduced by Wang et al. (2009) considering constrained contour curves and style curves [27]. To define general shape of clothes, contour curves such as silhouette curves and cross section curves were used whereas to generate detailed 2D cloth patterns on cloth surfaces style curves such as seam lines, notch lines and dart lines were used. This approach can be considered as an intuitive and appropriate way of designing complex cloth on a 3D mannequin. Further knowledge in fashion design and patternmaking is required to edit contour curves and style curves. Flexible shape control method was proposed Meng et al. (2012) for resizing 3D garments automatically while preserving shape of user-defined features on clothes [28]. Any kind of cloth modelling techniques can be used to generate 3D clothes are modelled on a reference human body. Safeguarding of cloth features can be highly influenced by inconsistent user inputs since feature curves are defined by users based on each cloth modelled for target human bodies. An automatic way of cloth transferring between characters with different body shapes was introduced by Brouet et al. (2012) [29]. Vertices are adjusted on 3D cloth in order to fit a cloth onto a different character and pattern extraction will be happened after fitting 3D cloth to a new character through this method. Sketch based interface for modelling 3D cloth on virtual characters was proposed by Turquin et al. (2007a) to model cloth on a 3D character based on stocks drawn by users [30]. This method can only be used to model simple style single layer clothes since information from user input stroke cannot be used to define complex cloth structure. An interactive tool was introduced by Umetani et al. (2011) for cloth design that enables bidirectional editing between 2D patterns and 3D cloth [31]. Real time physics-based simulation method is used to simultaneously update the topology of 2D cloth pattern and its corresponding 3D cloth piece by user input, in order to maintain synchronization between 2D and 3D. The method proposed through this method cannot be used for simulation of cloth dynamic behaviour subject to motion of 3D character as synchronization of 2D pattern with a 3D cloth on a static 3D mannequin cannot be performed through this method.

### B. Physical Based Virtual Clothing Methods

Triangular or rectangular grids with points of fixed mass at intersections are used to represent cloth models in these physical based virtual clothing methods. Two types of models such as energy-based method (Terzopoulos et al. 1987) and force-based method (Volino & Magnenat-Thalmann 2005) can be recognized in physical based virtual clothing methods [32] [33]. Total energy of cloth is calculated using some equations in energy-based model. The shape of cloth is derived using these equations by moving points to achieve minimum energy state. In force-based models, forces among each point are represented as differential equations. Positions of points at each time step are obtained by solving these differential equations using numerical integration. A method for the construction of shape of a cloth object has been proposed by Terzopoulos et al. (1987) [32]. Shape and motion of deformable materials has been described using elasticity theory in this approach. Simulation of dynamic behavior of objects can be implemented by fetching physical properties such as mass and damping in to physical simulation. General mechanical model for cloth simulation was introduced by Volino and Magnenat-Thalmann (2005) for cloth simulation [33]. Instead of mass-spring system, an accurate particle system for dynamic simulation is presented through this model. Three 2D coordinates have designed considering three mechanical properties such as weft, warp elongation and shear and these were used to describe a triangle face of cloth mesh. A simulation model for large deformations of textile was proposed by Volino et al. (2009) [36]. Simulation process has become simple through this model and it enables creation of nonlinear tensile behavior of textile with accuracy and robustness. An approach to simulate inextensible cloth in a collision-free condition subjected to a conservative force such as gravity was proposed by Chen and Tang (2010) [37]. Stretch resistance and compression resistance of a cloth is greater than its bending resistance according to this paper. Transformation of deformation process of an initial developable mesh surface to a final mesh surface through physical based simulation process has been proposed via this method.

In general physical based virtual clothing methods are used to produce behavior of flexible object that resembles cloth. Rate of resemblance will vary based on used physical based virtual clothing technique.

### C. Hybrid Virtual Clothing Methods

Hybrid virtual clothing methods have been developed by combining both geometrical based virtual clothing methods and physical based virtual clothing methods to compensate deficiencies in those methods to deliver a proper solution. Rudomin (1990) proposed a method to diminish the computational complexity of physical based virtual clothing methods [38]. He presented a mechanism which can be used to lower computational time of physical based virtual clothing method which was introduced by Terzopoulos et al. (1987) [32]. After a while a series of hybrid virtual clothing methods were proposed by Kunii and Gotoda (1990) [39]; Tsopelas (1991) [40]. Geometrically modelled fine wrinkle details were mapped onto a physically simulated cloth mesh using these methods. Texture based wrinkle modelling method was introduced by Hadap et al. (1999) for cloth simulation [41]. Deformation details will be generated through this method by considering basis as bump map which will be created by user on a physically simulated bristly cloth object. Kinematic method for generating wrinkles on cloth for computer generated (CG) characters was proposed by Cutler et al. (2005) [42]. Through this method they were able to find out that these similarities could be revealed only on tight fit cloth. Due to that it was impossible to generate shape detail on loose fit cloth using wrinkle database. A method to generate fine detailed folds on captured cloth model was proposed by Popa et al. (2009) [43]. Shape and position of wrinkles were captured from video footage through this method by considering basis as the distinguishing shape characteristics of wrinkle. Feng et al. (2010) introduced a method to provide high quality dynamic folds and wrinkle for cloth simulation by maintaining real time ability [44]. Geometrical based virtual clothing method was used to capture relationship between two different resolutions of mesh and transformation process was executed using this relationship. They developed an animation production pipeline which always starts with physical simulation of low-resolution mesh. Through this method they have evaluated collisions between each proxy bone instead of calculating collision between each mesh triangles of cloth and body model, when a character moves. Computational time taken to calculate collisions has reduced considerably even though there is a slight reduction in the accuracy of collision handling. Researchers have been succeeded in improving the efficiency, but a pre-simulation is required to obtain training data and it will be a time consuming process.

## VII. USAGE OF EASE ALLOWANCE AND DISTANCE EASE DISTRIBUTION IN PATTERN CONSTRUCTION

Ease has become a crucial feature in clothing industry in order to deliver appropriate garment fit to the wearer. Many researchers have identified a variety of factors that are having a considerable relation to the ease allowance of a garment block pattern construction. Rasband and Liechty (2006) have explained that design style, fabric physical and mechanical attributes, body shape, wearing occasion and personal preference are the main factors that impact the amount of ease needed [45]. Chen, Zeng, Happiette and Bruniaux (2008) have categorized the factors related to ease allowance into three; standard ease, dynamic ease and fabric ease [5]. According to Gill (2011), there are five contributing factors that determine the ease of a garment pattern such as Function, Comfort, Oversize, Fabric and Styling [46]. Huang, Mok, Kwok, & Au (2012) has identified the amount of clothing ease depends on the design, fabric used, functions of a garment, and even personal preference of customer [4]. Based on these, two type of eases were identified; wearing ease and design ease. As it can be seen, almost all the researchers have considered the shape of the human body and fabric properties to determine ease allowance [4][5][45][46], and these have been identified for having a strong and direct relationship with the distance ease distribution between body and garment [47][48]. Some of the researchers have identified that other than the body shape and fabric properties, the design of the garment, personal preferences and body movements also have a significant impact on the amount of ease allowance required. Still, there are questions left unanswered for the identification of the relationship between these factors and distance ease distribution when a garment is draped upon a human body.

Furthermore, many studies have used 3D body scanning data in order to evaluate the distance ease distribution [4][5][46][47][48][49]. Xu (2008) and Lage (2017) presented the impact on the distance ease between the body and garment by using a variety of fabrics, while uniformly changing the ease allowance. Both studies showed that the 3D distance eases at different body angles changed irregularly with the increase of garment sizes and the uniform changing of material mechanical properties, especially tensile strain [47][48]. Gill (2011) and Wang et al. (2006) have presented different mathematical models of ease distribution which plays an important role in the construction of basic garment patterns [46][50]. Chen et al. (2008) also presented a model to generate optimum ease allowance for the creation and manipulation of pattern construction [5]. Thomassey et al. (2013) identified a template to determine ease in 3D patterns in order to generate personalized garment patterns [49]. With the identification of distance ease distribution among body and garment, Huang et al. (2012) presented a model to flexibly distribute and accurately control ease in 3D patterns in order to convert those into two-dimensional [4].

Furthermore, studies have been conducted to evaluate the distance ease distribution at bust section, waist section, abdomen section, hip section, thigh, knee and crotch curve using the respective cross-sections of the body. Moreover, dresses, long-sleeve tops, jackets, trousers have been used for the evaluation of distance ease distribution.

There are still questions left unanswered for determining the garment fit for a given garment type without the usage of 3D technologies as well as for identifying the relationship between the distance ease distribution and wearer's personal fit preference.

## VIII. CONCLUSION

There are several methods and tools currently available to check the fitness of a garment with human body. Most of these techniques demand high computational power since they deliver results utilizing available 3D technologies. Therefore time required for these applications to deliver results is relatively high and high user intervention is needed to deliver the solution. Further users will access online marketplaces concurrently and it requires high speed delivery of results in order to control the network traffic. Thus the applicability of aforementioned 3D applications to an online marketplace would be comparatively low. Therefore there is a need for a lightweight solution which will support the decision making of the end user in order to select the properly fitted dress by considering human body properties and garment properties.